\documentclass[conference, letterpaper, 10pt, romanappendices]{IEEEtran}
\usepackage{graphics}
\usepackage{cite}
\usepackage[pdftex]{graphicx}
\usepackage{epstopdf}
\usepackage{epsfig}
\usepackage{latexsym}
\usepackage{amsfonts}
\usepackage{calc}
\usepackage{url}
\usepackage{enumerate}
\usepackage{color}
\usepackage[tbtags]{amsmath}
\usepackage{amssymb}
\usepackage{upref}
\usepackage{dsfont}
\usepackage{multirow}

\newtheorem{theorem}{Theorem}

\newtheorem{lemma}{Lemma}

\newtheorem{conjecture}{Conjecture}

\usepackage[left=.625in,top=.655in,right=.625in,bottom=.8in]{geometry}

\begin{document}
\title{Analysis of Overlapped Chunked Codes with Small Chunks over Line Networks}

\author{\IEEEauthorblockN{Anoosheh~Heidarzadeh and Amir H. Banihashemi\\}
\IEEEauthorblockA{Department of Systems and Computer Engineering, Carleton University, Ottawa, ON, Canada}}

\maketitle
\thispagestyle{empty}
\begin{abstract} To lower the complexity of network codes over packet line networks with arbitrary schedules, chunked codes (CC) and overlapped chunked codes (OCC) were proposed in earlier works. These codes have been previously analyzed for relatively large chunks. In this paper, we prove that for smaller chunks, CC and OCC asymptotically approach the capacity with an arbitrarily small but non-zero constant gap. We also show that unlike the case for large chunks, the larger is the overlap size, the better would be the tradeoff between the speed of convergence and the message or packet error rate. This implies that OCC are superior to CC for shorter chunks. Simulations consistent with the theoretical results are also presented, suggesting great potential for the application of OCC for multimedia transmission over packet networks.\end{abstract}



\section{Introduction}
There has recently been a surge of interest in application of coding for large-scale file sharing over packet networks. Network coding has been shown to generally reduce the expected file downloading time for various probabilistic or deterministic models of the flow transmission schedules. In practice, however, the accurate modeling of the schedule might be too complex and/or infeasible \cite{CWJ:2003,PFS:2005}. The problem of a practical code design is to achieve the capacity of the network under any arbitrary schedule with high probability.

Random linear network codes (a.k.a. dense codes) are known to achieve the capacity in this setting asymptotically (when the message length grows large), while having linear coding costs (the encoding/decoding algorithms at each node require a number of packet operations per message packet linear in the message length) \cite{MHL:2006,HB:2010}. This, however, impedes the application of dense codes for the transmission of large files. Thus the problem of code design from another perspective is to devise coding algorithms with relatively low complexity.


To overcome the computational inefficiency of dense codes, \emph{chunked codes} (CC) were proposed in \cite{MHL:2006}, and afterwards, a generalized version of chunked codes, referred to as \emph{overlapped chunked codes} (OCC), were independently proposed in \cite{HB:2010} and \cite{SZK:2009}.\footnote{The idea of overlapping chunks was proposed in \cite{HB:2010} and \cite{SZK:2009}, independently. Unlike \cite{HB:2010}, however, no theoretical result was presented in \cite{SZK:2009}.} These codes operate by dividing the original message into non-overlapping or overlapping chunks, respectively. Each node then randomly chooses a chunk at any time instant and transmits it by using a dense code.\footnote{There are a number of variations of codes based on chunks in the literature of network codes (see \cite{SZK:2009,LSS:2011}, and references therein). To the best of our knowledge, however, none of these codes have been provably shown to perform better than those in \cite{MHL:2006,HB:2010} over arbitrary schedules in the asymptotic regime.} Thus, CC and OCC require less complex coding operations as they apply coding on smaller chunks rather than the original message. (The coding costs of such codes are linear in the size of the chunks.)

Both CC and OCC are known to asymptotically achieve the capacity so long as the size of chunks (a.k.a. aperture size) is bounded below \cite{MHL:2006,HB:2010}. This lower bound has been shown to be an increasing function of the message length. Thus the aperture size cannot be reduced down to a constant in the message length. The coding algorithms, therefore, cannot be performed in linear time (with constant costs in the message length). This may hamper the use of such codes in practical applications under severe computational resource limitations.

Targeting the design of codes with smaller coding costs, the main contributions of this work are listed below:

\begin{itemize}
\item We prove that CC with small apertures provide a better tradeoff between the speed of convergence and the message or packet error rate in comparison to what was previously thought, based on the results of \cite{MHL:2006}.

\item As part of our machinery, we generalize a recently proposed conjecture in \cite{SB:2008} on the rank property of a special class of random matrices with overlapping bands.\footnote{The validity of our conjecture is supported via simulations, not included in this paper due to space limitations, but a formal proof is still unknown.}


\item We show that unlike the case for CC with small apertures, a martingale argument does not lead to a tight analysis of OCC with small apertures. This is due to higher levels of dependency between recoverability of message packets in the case of overlapping chunks. We thus adopt a new methodology studying two extremal types of dependency and give a tight analysis for each type.

\item We prove that for small apertures: (i) CC and OCC approach the capacity with an arbitrarily small but positive constant gap; (ii) OCC with larger overlaps are superior in terms of the tradeoff between the speed of convergence and the message or packet error rate. This is the opposite of the trend for sufficiently large apertures, where OCC with smaller overlaps are known to be superior \cite{HB:2010}.

\item We show the advantage of OCC over CC via simulations. For example, when compared to a CC with similar coding costs, the application of an OCC can decrease the downloading time of a $1$MB file from a file server $4$ hops away by about $15\%$ to $30\%$ depending on the target packet error rate and the chunk size.
\end{itemize}


For more details, including the proofs, please see~\cite{HBJ:2011}. 

\section{Model and Problem Statement}\label{ModelProblemStatement}
We focus on the unicast problem over line networks. We follow the terminology and notations of \cite{HB:2010}. Consider a line network of length $l$, where one \emph{source} node (which is given $k$ message packets, where the packets are strings of bits) is connected to one \emph{sink} node (which demands the $k$ message packets) through $l$ links in tandem. The rest of the $l-1$ nodes are called \emph{interior}. The links are modeled by erasure channels. The erasure probabilities are arbitrary and may vary over time. The packets might be faced with arbitrary delays. The connectivity graph specifying the flow of the successful transmissions over time is called a \emph{schedule}.

The schedule describes paths over which packet transmissions are successful \cite{HB:2010}. The min-cut capacity of the schedule equals the number of edge-disjoint paths within the schedule starting from the source node and ending at the sink node. Since we are interested in the analysis of coding schemes over arbitrary schedules of a given capacity $n$, we study the worst-case schedules in which the probability of success is the smallest possible (subject to having the least possible number of successful packets providing the given capacity). We suppose that there are precisely $n$ packets successfully sent and/or received by each node; and there exists one and only one packet departure from each interior node between any two consecutive packet arrivals at the node.

The goal is to analyze and compare our codes of interest over line networks of length $l$ with worst-case schedules of capacity $(1+\lambda)k$, for any arbitrarily small but constant $\lambda>0$, as $k$ goes to infinity. We specifically focus on the tradeoff between the speed of convergence and the probability of decoding failure (the message error rate) or the expected fraction of undecodable message packets (the packet error rate).


\section{CC with Small Apertures}\label{CCSA}
A CC operates by partitioning the $k$ message packets into $q$ non-overlapping chunks, each of aperture size $\alpha=k/q$.

Each node randomly chooses a chunk at each transmission time and transmits a random linear combination (over binary field) of the previously received packets pertaining to the chosen chunk. To recover each chunk, the sink node has to solve a linear system of equations for $\alpha$ message packets. It is easy to see that the coding cost per network node is $O(\alpha)$.\footnote{The coding cost per network node is the number of packet operations per message packet required for performing coding algorithms at each node.}


For every $\omega\in[q]$, the packets pertaining to the chunk $\omega$, are called \emph{$\omega$-packets}. Each set of packets with linearly independent local (global) encoding vectors is called \emph{dense} (\emph{innovative}). The packets belonging to a dense (innovative) set are called \emph{dense} (\emph{innovative}). The decoding is successful if every chunk is decodable, and for every $\omega$, chunk $\omega$ is decodable if there exist $\alpha$ innovative $\omega$-packets at the sink node.

Now suppose that a CC with $q$ chunks of size $\alpha$ is applied to $k$ message packets at the source node of a network of length $l$ with any worst-case schedule of capacity $(1+\lambda)k$.

The chunks are randomly scheduled over time, and hence the capacity of the flow by the $\omega$-packets (the capacity of the schedule restricted to the flow of $\omega$-packets), for every $\omega$, is a random variable. This capacity can be lower bounded with high probability (w.h.p.) by using a probabilistic counting technique, so long as the aperture size is bounded above (b.a.).

\begin{lemma}\label{FlowChunkedCodeIIWorstCase} For a CC with $q$ chunks, over a network of length $l$ with any worst-case schedule of capacity $n$, for every $\omega$, the $\omega$-packets fail to form a flow of capacity not larger than \begin{equation}\label{FlowChunkedCodeIIEqWorstCase}\left(1-O\left(\left(l^3({q}/{n})\ln({ln}/{\epsilon})\right)^{1/3}\right)\right).(n/q)\end{equation} with probability (w.p.) bounded above by (b.a.b.) $\epsilon$, so long as \begin{equation}\label{ConditionIWorstCase}l^3q\ln\frac{ln}{\epsilon}=o\left(n\right).\footnote{By replacing $q$ with $k/\alpha$, the condition~\eqref{ConditionIWorstCase} can be interpreted as a lower bound on the aperture size $\alpha$.}\end{equation}\end{lemma}

The result of Lemma~\ref{FlowChunkedCodeIIWorstCase} is tighter than \cite[Lemma~3]{HB:2010}, which itself is tighter than \cite[Theorem~4.1]{MHL:2006}, and hence any succeeding result in this paper would be tighter than its counterpart in \cite{MHL:2006}.


By using the lower bound on the capacity of the flow by $\omega$-packets and applying \cite[Corollary~1]{HB:2010}, for every $\omega$, the number of dense $\omega$-packets are lower bounded w.h.p. Next, by using the lower bound on the number of dense $\omega$-packets and applying \cite[Lemma~2]{HB:2010}, for every $\omega$, the probability that the sink node receives $\alpha$ innovative $\omega$-packets is given a lower bound.

These results however are not applicable to the underlying problem in this paper as the aperture sizes of our interest are expected to be smaller than the lower bound given in Lemma~\ref{FlowChunkedCodeIIWorstCase}. 

We, instead, fix a particular chunk, say $\omega$, and lower bound the capacity of the schedule restricted to the flow by $\omega$-packets. We are then able to lower bound on the probability of receiving a collection of $\alpha$ innovative $\omega$-packets by the sink node.

\begin{lemma}\label{FailureCCNoUnionBound}For any $\epsilon>0$, and $\lambda>0$, applying a CC with aperture size $\alpha$ over a network of length $l$ with any worst-case schedule of capacity $(1+\lambda)k$, the probability of a given chunk $\omega$ to be undecodable is b.a.b. ${\epsilon}$, so long as \begin{equation}\label{FailureCCNoUnionBoundEq}\alpha\leq \varphi-l\log(l\varphi/\dot{\epsilon})-\log(1/{\epsilon})-l-1,\end{equation} where \begin{equation}\label{CCFlowNoUnionBoundEq}\varphi= \left(1-O\left(\left((l^3/\mu)\ln(l\mu/{\epsilon})\right)^{1/3}\right)\right)\cdot \mu,\end{equation} $\mu=(1+\lambda)\alpha$, and $\dot{\epsilon}=\epsilon/2$.\end{lemma}


Substituting~\eqref{CCFlowNoUnionBoundEq} into~\eqref{FailureCCNoUnionBoundEq}, we obtain \begin{equation}\label{ApertureSize1}\alpha= \Omega\left(({l^3}/{\lambda^3})\ln({l}/{\lambda\epsilon})\right).\end{equation}

By applying a union bound on the result of Lemma~\ref{FailureCCNoUnionBound} together with condition~\eqref{ApertureSize1}, the following is immediate.

\begin{theorem}\label{CCCAWC} For any $\epsilon>0$, when ${\epsilon}$ goes to $0$ sufficiently fast, as $k$ goes to infinity,\footnote{We say $\epsilon$ goes to $0$ ``sufficiently fast,'' if $\epsilon q$ vanishes as $k$ goes to infinity.} applying a CC with $\alpha=\Omega(({l^3}/{\lambda^3})\ln({l}/{\lambda\epsilon}))$, to $k$ message packets over a network of length $l$ with any worst-case schedule of capacity $(1+\lambda)k$, not all the chunks are decodable w.p. b.a.b. ${\epsilon} q$.\end{theorem}


Now, let us assume larger values of $\epsilon$ up to a constant. By constructing a martingale sequence over the number of chunks that are not decodable while exposing chunks one at a time and applying Azuma's inequality, see \cite[Chapter~7]{AS:2008}, the actual fraction of chunks which fail to be decoded can be shown to be tightly concentrated around its expectation.



\begin{theorem}\label{DeviationExpSpWC}For any $\epsilon>0$, and $\lambda>0$, for a CC with $\alpha=\Omega\left(({l^3}/{\lambda^3})\ln({l}/{\lambda\epsilon})\right)$, over a network of length $l$ with any worst-case schedule of capacity $(1+\lambda)k$, for any $\gamma_a>0$, the fraction of undecodable chunks deviates farther than $\gamma_a$ from ${\epsilon}$, w.p. b.a.b. $e^{-ck}$, for some constant $c=O(\gamma_a^2\epsilon^2/\alpha^2)$.\footnote{Note that, however, for relatively large values of $\epsilon$, the expected fraction of unrecoverable message packets is bounded away from zero. Thus, a CC, alone, does not recover all the message packets. One solution is to devise a proper precoding scheme. The issue of the precode design is beyond the scope of this paper. Readers are referred to \cite{MHL:2006} for more information.}\end{theorem}


\section{OCC with Small Apertures}\label{OCCSA}
An OCC splits the message packets of size $k$ into $q$ chunks of size $\alpha$, so that any two contiguous chunks overlap by $\gamma=\alpha/\tau_e$ message packets (in an end-around fashion), where $\tau_e=\tau/(\tau-1)$, for any constant integer divisor $\tau$ of $\alpha$.

The encoding is similar to that in CC, and the encoding cost per network node is $O(\alpha)$. Unlike CC, the decoding is similar to that in dense codes, i.e., to recover the chunks all together, the sink node has to solve a system of linear equations for $k$ message packets.\footnote{In prior related work, excluding \cite{HB:2010}, the chunks are to be decoded in isolation. However, by performing the decoding algorithm on the set of all the chunks simultaneously, not all the chunks need to be recoverable in isolation when the chunks are overlapping. Thus a smaller number of packets at the sink node is sufficient to ensure successful decoding with a given probability. This, however, may come at the cost of increasing the memory requirements, yet studying the latter tradeoff is beyond the scope of this paper.} The matrix of the coefficients associated to such a system is a banded matrix of bandwidth $\alpha$ (the non-zero elements of each row lie within a band of size $\alpha$). Such a system can be solved by using $O(\alpha)$ row (packet) operations \cite{SB:2008}, and the decoding cost at the sink node is $O(\alpha)$.



Unlike CC, in OCC, for a given $\omega$, an innovative collection of $\omega$-packets might not be innovative to the set of all the packets. This arises from the fact that any given chunk shares a few message packets with a few other chunks. The decoding is successful if the chunks are all decodable as a set, i.e., there exist $k$ innovative packets (regardless of the number of innovative $\omega$-packets for some $\omega$) at the sink node.

Suppose an OCC with $q$ $(=k\tau/\alpha)$ chunks, each of size $\alpha$, and overlap $\gamma=\alpha/\tau_e$, applied to $k$ message packets at the source node of a network of length $l$ with any worst-case schedule of capacity $(1+\lambda)k$.

The same approach used for the analysis of CC with small apertures is not applicable to OCC with small apertures. In particular, unlike CC, in OCC, a martingale argument alone does not give a tight upper bound on the fraction of unrecoverable message packets. This arises from the fact that in OCC, the chunks are to be decoded together. 


In the following, we provide a sketch of our analysis, starting with giving a useful conjecture on the rank property of two classes of banded random binary matrices.

Let $n,k,\alpha$ and $\gamma$ be integers ($k\leq n$, $\gamma<\alpha$), so that $\alpha-\gamma$ is a divisor of $k$. Let $\chi$ be $k/(\alpha-\gamma)$, and $I$ be the set of integers $[k]$. We divide $I$ into $\chi$ subsets $I_i$'s, for all $i\in[\chi]$, where $I_i$ (the $i$th \emph{aperture} of size $\alpha$) is the set of $\alpha$ contiguous integers in $I$ in an end-around fashion, starting from $(i-1)(\alpha-\gamma)+1$.

We construct an $n\times k$ matrix as follows: (i) for each row, an index, say $i$, is randomly chosen from the set of integers $[\chi]$, and (ii) the row's entries indexed by the $i$th aperture are independently and uniformly chosen from the binary field, and the rest of the entries are set to zero. We call such a matrix a $(\gamma,\alpha)$ irregular symmetric banded matrix of size $n\times k$. Now, consider a similar construction, but when $\alpha-\gamma$ is a divisor of $k-\gamma$ (not $k$), and $\chi$ is $(k-\gamma)/(\alpha-\gamma)$ (not $k/(\alpha-\gamma)$). The resulting matrix in this case is called a $(\gamma,\alpha)$ irregular asymmetric banded matrix of size $n\times k$. Further, consider a matrix constructed as above, except that in part (i), each index in $[\chi]$ is assigned to $n/\chi$ rows ($\chi$ has to be a divisor of $n$). We call such a matrix a $(\gamma,\alpha)$ ``regular'' symmetric or asymmetric banded matrix of size $n\times k$.

\begin{conjecture}\label{RankAperture}Let $n,k,\alpha$ and $\gamma$ be integers ($k\leq n$, $\gamma<\alpha$). Let $M$ be a $(\gamma,\alpha)$ (regular/irregular) symmetric or asymmetric banded random matrix of size $n\times k$. For any $\epsilon>0$, and for sufficiently large $k$, $\Pr[r(M)<k]\leq\epsilon$, so long as $k\leq n-\log(1/\epsilon)$, and $\gamma\geq 2\sqrt{k}$, or $\gamma\geq \tau_e\tau\sqrt{k}$, respectively, where $r(M)$ is the rank of the matrix $M$ over the binary field, $\gamma=\alpha/\tau_e$, and $\tau_e=\tau/(\tau-1)$, for any constant divisor $\tau$ of $\alpha$.\footnote{We briefly highlight the differences between Conjecture~\ref{RankAperture} and Conjecture~4.2 of \cite{SB:2008}: (i) the latter considers a subclass of symmetric banded random binary matrices, yet the former considers two more general classes of regular/irregular symmetric and asymmetric banded random binary matrices, and (ii) unlike the latter, for a given aperture size, the overlap size in the former is not restricted to one particular value.}\end{conjecture}

Now consider the matrix of the global encoding vectors of the received packets at the sink node, $Q$, also referred to as the ``decoding matrix.'' Let $Q'$ be $Q$ restricted to its dense rows (global encoding vectors of the dense packets). Let $\chi$ be an integer sufficiently smaller than the number of chunks. Fix a particular set of $\chi$ contiguous chunks (we will specify the precise choice of $\chi$ later). Focus on the set of dense packets pertaining to the given set of chunks. We first lower bound the probability that the set of rows pertaining to these chunks in $Q'$ includes a $(\gamma,\alpha)$ regular asymmetric banded matrix with $\chi(\alpha-\gamma)+\gamma$ columns (the number of distinct message packets in $\chi$ contiguous chunks) and a sufficiently large number of rows. By using Conjecture~\ref{RankAperture}, we next upper bound the probability that such a set fails to be decodable. Studying all such sets, the fraction of recoverable message packets can be lower bounded (all the message packets in a chunk belonging to a decodable set of chunks are recoverable).\footnote{It is worth noting that our analysis is sub-optimal in the sense that there might be some recoverable message packets that we declare as unrecoverable. This is because the decoding is performed on the set of all the chunks together, not on the subsets of chunks in isolation.}

We formalize the above process as follows. We call each set of $\chi$ contiguous chunks, in an end-around fashion, a \emph{hyperchunk}. We call each (disjoint) set of $\alpha/\tau$ contiguous message packets a \emph{block}. We say that a given hyperchunk is not decodable (a \emph{bad hyperchunk}) if it fails to be decoded \emph{in isolation}. We also say that a given block is not recoverable (a \emph{bad block}) if it does not belong to any decodable hyperchunk.


Followed by lower bounding the capacity of the flow by the packets pertaining to a given hyperchunk, Lemma~1 of \cite{HB:2010} serves to bound the probability of receiving an innovative collection of $\chi k/q$ packets belonging to this hyperchunk.

\begin{lemma}\label{FailureOCCNoUnionBound} For any $\epsilon>0$, and $\lambda>0$, applying an OCC with aperture size $\alpha$, and overlap parameter $\tau$, over a network of length $l$ with any worst-case schedule of capacity $(1+\lambda)k$, the probability of a given hyperchunk of size $\chi$ to be bad is b.a.b. ${\epsilon}$, so long as \begin{equation}\label{BigIneq} r\alpha\leq \chi\varphi-\chi l \log(l\varphi\chi/\dot{\epsilon})-\log(1/\dot{\epsilon})-\chi l,\end{equation} given that $\gamma\geq \tau_e\tau\sqrt{r\alpha}$, where \begin{equation}\label{CapacityOCCP}\varphi=\left(1-O\left(\left((l^3/\mu)\ln(l\mu\chi/{\epsilon})\right)^{1/3}\right)\right)\cdot \mu,\end{equation} $\mu=(1+\lambda)\alpha/\tau$, and $r=(\chi-1)/\tau+1$.\end{lemma}


Substituting~\eqref{CapacityOCCP} into~\eqref{BigIneq}, we obtain \begin{equation}\label{ApertureSizeII}\alpha=\Omega\left((l^3/\lambda^3)\tau\ln\left((l/\lambda\epsilon)\tau\right)\right),\end{equation} by choosing $\chi$ to be an arbitrary constant integer sufficiently larger than $(\tau-1)/\lambda$.

Thus the probability that a given hyperchunk is bad is b.a.b. ${\epsilon}$, and the expected fraction of bad hyperchunks is upper bounded by ${\epsilon}$. By using a martingale argument over the hyperchunks, one can show that for any $\gamma_a>0$, the fraction of undecodable hyperchunks deviates farther than $\gamma_a$ from ${\epsilon}$, w.p. b.a.b. $e^{-ck}$, for some constant $c=O((\gamma_a^2{\epsilon}^2/\alpha^2)(\lambda\tau))$.



Now, the problem is to upper bound the fraction of bad blocks. This fraction however depends on the ordering of bad hyperchunks. It should be clear that among different orderings of bad hyperchunks, the one in which all the bad hyperchunks are adjacent results in the largest fraction of bad blocks. Since the expected fraction of bad hyperchunks is ${\epsilon}$, the expected fraction of bad blocks is therefore upper bounded by ${\epsilon}$.



This, however, is not a tight bound in that the probability of a large number of bad hyperchunks being adjacent is very small. To give a tighter bound, various orderings of bad hyperchunks (or bad blocks) need to be studied. The ordering of bad hyperchunks is random, and so is the fraction of bad blocks. However the structure of the dependency between the hyperchunks (or blocks) is not easy to formulate.



We, instead, analyze two extremal types of dependency structures of hyperchunks as defined below. Let $I$ be the set of integers in $[q]$. For all $i\in I$, let $\mathcal{G}_i$ ($\mathcal{B}_i$) be the set of indices of the message packets in the $i$th hyperchunk (block). We use the same notation $\mathcal{G}_i$ ($\mathcal{B}_i$) to refer to the $i$th hyperchunk (block) unless there is a danger of confusion. We, further, let $G_i$ ($B_i$) be the event that $\mathcal{G}_i$ ($\mathcal{B}_i$) is not decodable (recoverable). For $i\in I$, let $N_{\mathcal{G}}(i)$ ($N_{\mathcal{B}}(i)$) be an ordered set (in an increasing cyclic order) of indices of hyperchunks that overlap with the $i$th hyperchunk (block), and $I_i$ be an arbitrary subset of $I\setminus\{i\}$. The first (second) type is the one that the occurrence of any subset of $G_j$'s, for $j\in N_{\mathcal{G}}(i)$ ($j\neq i$), increases (decreases) the probability that $G_i$ occurs. In both types, for $I_i\cap N_{\mathcal{G}}(i)=\emptyset$, the occurrence of any subset of $G_j$'s decreases the probability of occurrence of $G_i$. We refer to the first (second) type as the \emph{dependency with} \emph{positively} (\emph{negatively}) \emph{dependent neighborhoods}.

We upper bound (i) the probability that not all the blocks are recoverable, and (ii) the expected fraction of unrecoverable blocks. Such bounds are ``outer'' upper bounds for the class of dependency structures of the underlying type in that they hold for any dependency structure in the class. We say that an outer bound is ``tight'' over the class of dependency structures of a given type, if, in the limit of interest (as $k$ tends to infinity, the expected fraction of bad blocks vanishes sufficiently fast, or does not), it is tight for any worst-case structure in the class.


To give tight outer upper bounds for each type, we study the worst case, i.e., providing that any arbitrary subset of hyperchunks is not decodable, the conditional probability of undecodability of any given hyperchunk is the largest possible. These (tight) bounds indicate an interval that for any possible type of dependency, a tight outer upper bound lies within. The followings summarize our analytical results.




\begin{theorem}\label{OCCCAWC}For any $\epsilon>0$, when ${\epsilon}$ goes to $0$ sufficiently fast, as $k$ goes to infinity, and $\lambda>0$, applying an OCC with aperture size $\alpha=\Omega(({l^3}/{\lambda^3})\tau\ln(({l}/{\lambda\epsilon})\tau))$, and overlap parameter $\tau$, over a network of length $l$ with any worst-case schedule of capacity $(1+\lambda)k$, for any type of dependency between hyperchunks, there exists a tight outer upper bound on the probability that not all the blocks are recoverable. This bound is between ${\epsilon^{\chi+\tau-1}}q$ and ${\epsilon^2}q$, where $\chi$ is an arbitrary constant integer sufficiently larger than $(\tau-1)/\lambda$.\end{theorem}

\begin{theorem}\label{DeviationExpSpWCOCC} For any $\epsilon>0$, and $\lambda>0$, applying an OCC with aperture size $\alpha=\Omega\left(({l^3}/{\lambda^3})\tau\ln(({l}/{\lambda\epsilon})\tau)\right)$, and overlap parameter $\tau$, over a network of length $l$ with any worst-case schedule of capacity $(1+\lambda)k$, for any type of dependency between hyperchunks, there exists a tight outer upper bound on the expected fraction of unrecoverable blocks. This bound is between ${\epsilon^{\chi+\tau-1}}$ and $\epsilon^2$, where $\chi$ is an arbitrary constant integer sufficiently larger than $(\tau-1)/\lambda$. \end{theorem}

\section{Comparison Results}\label{Comparison}


By comparing Theorems~\ref{CCCAWC} and~\ref{OCCCAWC}, or Theorems~\ref{DeviationExpSpWC} and~\ref{DeviationExpSpWCOCC}, regarding the tradeoff between the message or packet error rate and the speed of convergence, the followings can be shown.


Let the $k$ message packets be divided into $\tau q$ chunks of size $\alpha$ ($=k/q$). We say that a code with $\alpha=\Omega((l/\lambda)^3\tau\ln((l/\lambda\epsilon)\tau))$ has ``relatively'' or ``very'' small aperture, if $\epsilon \tau q$ goes to zero or not, as $k$ goes to infinity.\footnote{By definition, the overlap parameter $\tau$ is smaller than or equal to the aperture size $\alpha$. The two lower bounds on $\alpha$, however, are increasing functions of $\tau$. Thus, for a given $\alpha$ to fall in the category of the relatively/very small chunks, $\tau$ needs to be bounded from above by a value smaller than $\alpha$. Therefore, it might not always be possible to make $\tau$ as large as $\alpha$.}

\begin{theorem}\label{Cor:3}For sufficiently small given message error rate and for relatively small apertures, the larger is the overlap, the larger is the speed of convergence.\end{theorem}

\begin{theorem}\label{Cor:5}For sufficiently small given packet error rate and for very small apertures, the larger is the overlap, the larger is the speed of convergence.\end{theorem}

\section{Simulation Results}\label{SimulationResults}
We compare CC and OCC in terms of the tradeoff between the overhead per message $\lambda:=(n-k)/k$, called ``overhead'' for brevity, and the message or packet error rate. The variables in this comparison are the message, aperture and overlap sizes.

We consider line networks of length $4$. The networks are simulated with randomly generated worst-case schedules of capacity $n=(1+\lambda)k$, for some $0\leq \lambda\leq 3$, and with the message sizes $k=64,256$. For each $k$, we consider the aperture sizes $\alpha=k/2,k/4$, and the overlap sizes $\gamma=\alpha(\tau-1)/\tau$, for the overlap parameters $\tau=1$ (i.e., CC), $2,$ or $4$. We are interested in the message and packet error rate, each as a function of the overhead for a given aperture size (given coding costs). For each $n,k,\alpha$ and $\gamma$, each coding scheme is applied to the schedules until $1000$ decoding failures occur.

\begin{figure}[t]
\centering
\includegraphics[width=2.75 in]{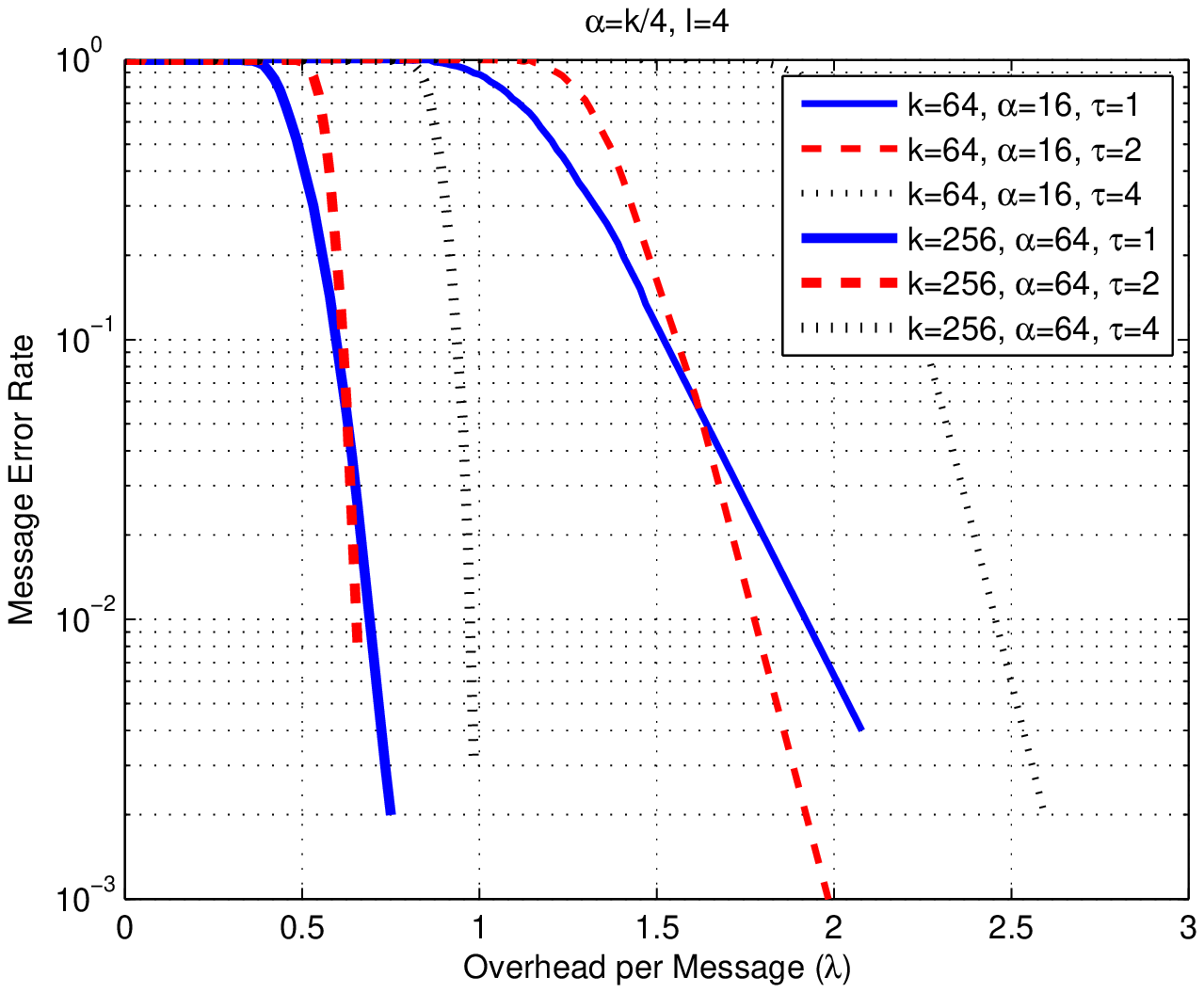}
\vspace{-.15 in}
\caption{CC and OCC: more computational efficiency.}\label{Fig62}
\end{figure}

\begin{figure}[t]
\centering
\includegraphics[width=2.75 in]{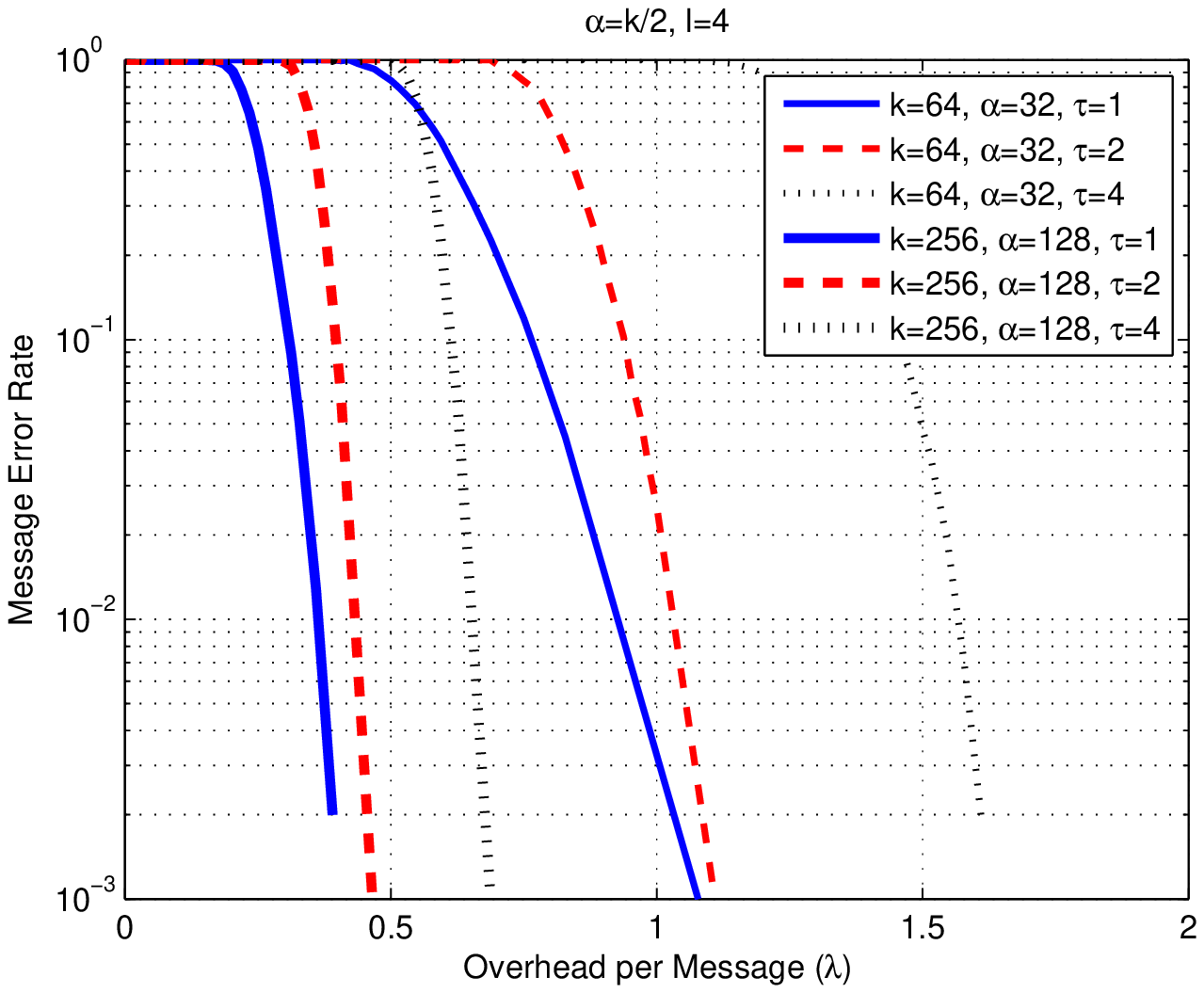}
\vspace{-.15 in}
\caption{CC and OCC: less computational efficiency.}\label{Fig61}
\end{figure}

\begin{figure}[t]
\centering
\includegraphics[width=2.75 in]{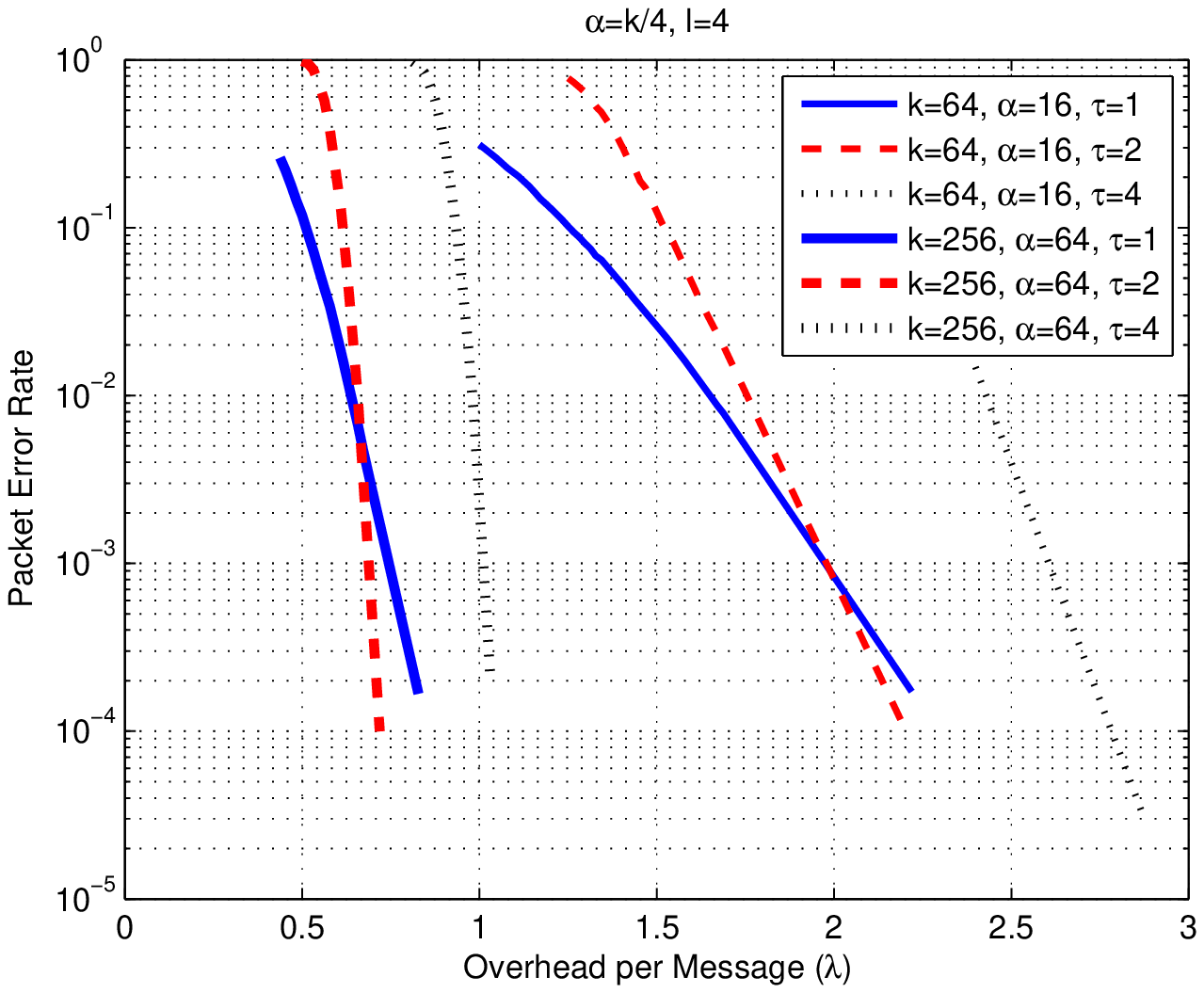}
\vspace{-.15 in}
\caption{CC and OCC: more computational efficiency.}\label{Gig62}
\end{figure}

\begin{figure}[t]
\centering
\includegraphics[width=2.75 in]{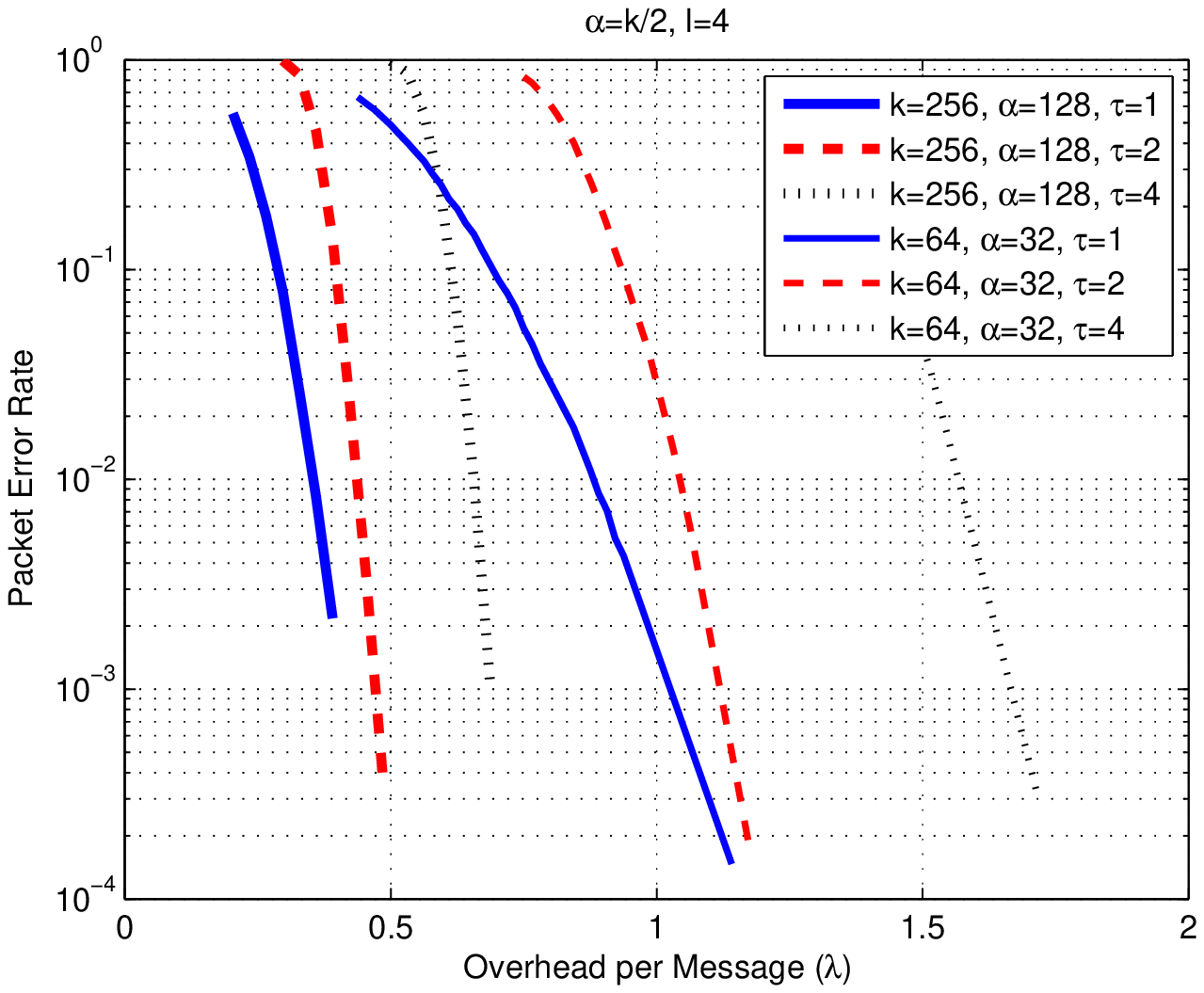}
\vspace{-.15 in}
\caption{CC and OCC: less computational efficiency.}\label{Gig61}
\end{figure}


Figure~\ref{Fig62} depicts the message error rate (MER) vs. the overhead $\lambda$ for the cases where the aperture size $\alpha=k/4$. Figure~\ref{Fig61} depicts the same scenarios only for the larger aperture size of $\alpha=k/2$. As the coding complexity is a linear function of $\alpha$, the cases considered in Figure~\ref{Fig61} are computationally less efficient than the corresponding cases in Figure~\ref{Fig62}.

Figure~\ref{Fig62} shows that for given number of message packets $k$, and a fixed coding complexity (fixed $\alpha$), if the target MER is sufficiently small, then OCC with larger overlap (larger $\tau$) outperforms OCC with smaller overlap (including CC), in terms of the convergence speed. For example, one can see that for $k=64$, and $\alpha=16$, for MERs below about $5\times 10^{-2}$, OCC with $\tau=2$ requires a smaller overhead compared to CC ($\tau=1$).

Comparison of Figures~\ref{Fig62} and~\ref{Fig61} reveals that the MER below which OCC outperforms CC is a decreasing function of the aperture size. So, the advantage of OCC over CC is more pronounced for smaller aperture sizes and lower target MERs.

Figures~\ref{Gig62} and~\ref{Gig61} show the packet error rate (PER) of the scenarios identical to those in Figures~\ref{Fig62} and~\ref{Fig61}, respectively, and the trends for PER results are similar to those of MER.

\emph{Example:} Consider downloading a 1MB file from a file server $4$ hops away ($l=4$). Suppose that packets of length 4KB are used for the transmission. This implies that the number of message packets $k=256$. Consider a target PER of $10^{-4}$, and two possible transmission scenarios: (a) using a CC with $\alpha = 64$, and (b) using an OCC with $\alpha = 64$ and $\tau = 2$ ($\gamma = 32$). From Figure~\ref{Gig62}, one can see that the overhead $\lambda$ for the two scenarios is about 0.85 and 0.7, respectively. This implies that downloading the file by scenario (b) is about $17\%$ faster than scenario (a). The difference will be even more for smaller target PERs and smaller aperture sizes.

\bibliographystyle{IEEEtran}
\bibliography{IEEEabrv,Refs}

\end{document}